\documentclass[
   ,final            % use final for the camera ready runs
  ]
  {aipproc}

\layoutstyle{6x9}

\usepackage{aas_macros}
\usepackage{multicol}

\begin{document}

\title{The Space Density of Compton-thick AGN}


\author{Ezequiel Treister}{
  address={Institute for Astronomy, University of Hawaii}
}

\author{Meg Urry}{
  address={Yale University}
}

\author{Carolin Cardamone}{
  address={Yale University}
}

\author{Shanil Virani}{
  address={Yale University}
}

\author{Kevin Schawinski}{
  address={Yale University}
%  ,altaddress={<author1 address>} % additional visiting address
}

%\author{David Sanders}{
%  address={Institute for Astronomy, University of Hawaii}
%}

\author{Eric Gawiser}{
  address={Rutgers University}
}

\begin{abstract}
We constrain the number density and evolution of Compton-thick Active Galactic Nuclei (AGN), and their
contribution to the extragalactic X-ray background. In the local Universe we use the wide area surveys from 
the {\it Swift} and {\it INTEGRAL} satellites, while for high redshifts we explore candidate selections based on 
mid-IR parameters. We present the properties of a sample of 211 heavily-obscured AGN candidates in the 
Extended {\it Chandra} Deep Field-South (ECDF-S) selecting objects with $f_{24\mu m}$/$f_R$$>$1000 
and $R$-$K$$>$4.5.The X-ray to mid-IR ratios for these sources are significantly larger than that of star-forming 
galaxies and $\sim$2 orders of magnitude smaller than for the general AGN population, suggesting column 
densities of $N_H$$>$5$\times$10$^{24}$~cm$^{-2}$. The space density of CT AGN at $z$$\sim$2 derived from 
these observations is  $\sim$10$^{-5}$Mpc$^{-3}$, finding a strong evolution in the number of  
$L_X$$>$10$^{44}$~erg/s sources from $z$=1.5 to 2.5.
\end{abstract}

\maketitle

%%%%%%%%%%%%%%%%%%%%%%%%%%%%%%%%%%%%%%%%%%%%
%% MAINMATTER
%%%%%%%%%%%%%%%%%%%%%%%%%%%%%%%%%%%%%%%%%%%%

\section{Compton Thick AGN with INTEGRAL and Swift}

The most obscured AGN known are those in which the neutral hydrogen
column density ($N_H$) in the line of sight is higher than the inverse
Thomson cross section, $N_H$$\simeq$1.5$\times$10$^{24}$~cm$^{-2}$. These are the so-called
Compton-thick (CT) AGN. Contrary to the situation for less obscured sources, not 
much is known about the number density of CT AGN. Thanks to the deep {\it Chandra} and
{\it XMM-Newton} surveys it is now clear that the fraction of moderately obscured, Compton-thin, AGN is on 
average $\sim$3/4 of all AGN, and is higher at lower luminosities \citep{ueda03,treister05b} and higher 
redshifts \citep{treister06b}, but there are no comparable constraints on the number of CT AGN.
One of the best ways to find CT AGN is by observing at high energies, namely E$>$10~keV. One clear 
advantage of these high-energy observations is that photoelectric absorption has minimal effects, so even CT AGN 
can be easily detected. 

Using the IBIS coded-mask telescope, {\it INTEGRAL} surveyed $\sim$80\% of the sky down to a flux of 5 mCrab in the
17-60~keV band. The catalog of \citet{krivonos07} reports the properties of 130 sources detected in these all-sky observations 
and classified as AGN. Five of the 130 AGN are CT AGN. Similarly, \citet{tueller08} presented a catalog of 103 AGN 
detected in an all-sky survey with the {\it Swift/BAT} telescope.  Excluding blazars, BL Lac and low galactic latitude 
observations, we obtained a sample of 89 sources, where only one source remains unidentified.  In the \citet{tueller08} 
catalog there are five AGN with estimated $N_H$ greater than 10$^{24}$~cm$^{-2}$.  Figure~\ref{logn_s} shows the 
cumulative number counts of AGN, with CT sources shown separately, as a function of hard X-ray flux. In order to 
avoid the necessity of specifying a standard spectrum to convert fluxes to different energy bands, we show 
the {\it INTEGRAL} and {\it Swift} sources separately, but note that a good agreement (within $\sim$40\%) in the 
normalization between the two distributions exist if a standard band conversion is assumed. At these high fluxes 
the slope of the $\log$~N-$\log$~S is Euclidean, implying an uniform spatial distribution, as expected given the 
low redshifts of these sources. We also compare with the distribution predicted by the AGN population synthesis 
model with which \citet{treister05b} fit the XRB, and find in general good agreement in slope and normalization.

%\vspace{-1.3cm}
\begin{ltxfigure}[h!]
\begin{minipage}{3.6in}
%\begin{center}
\includegraphics[width=\textwidth]{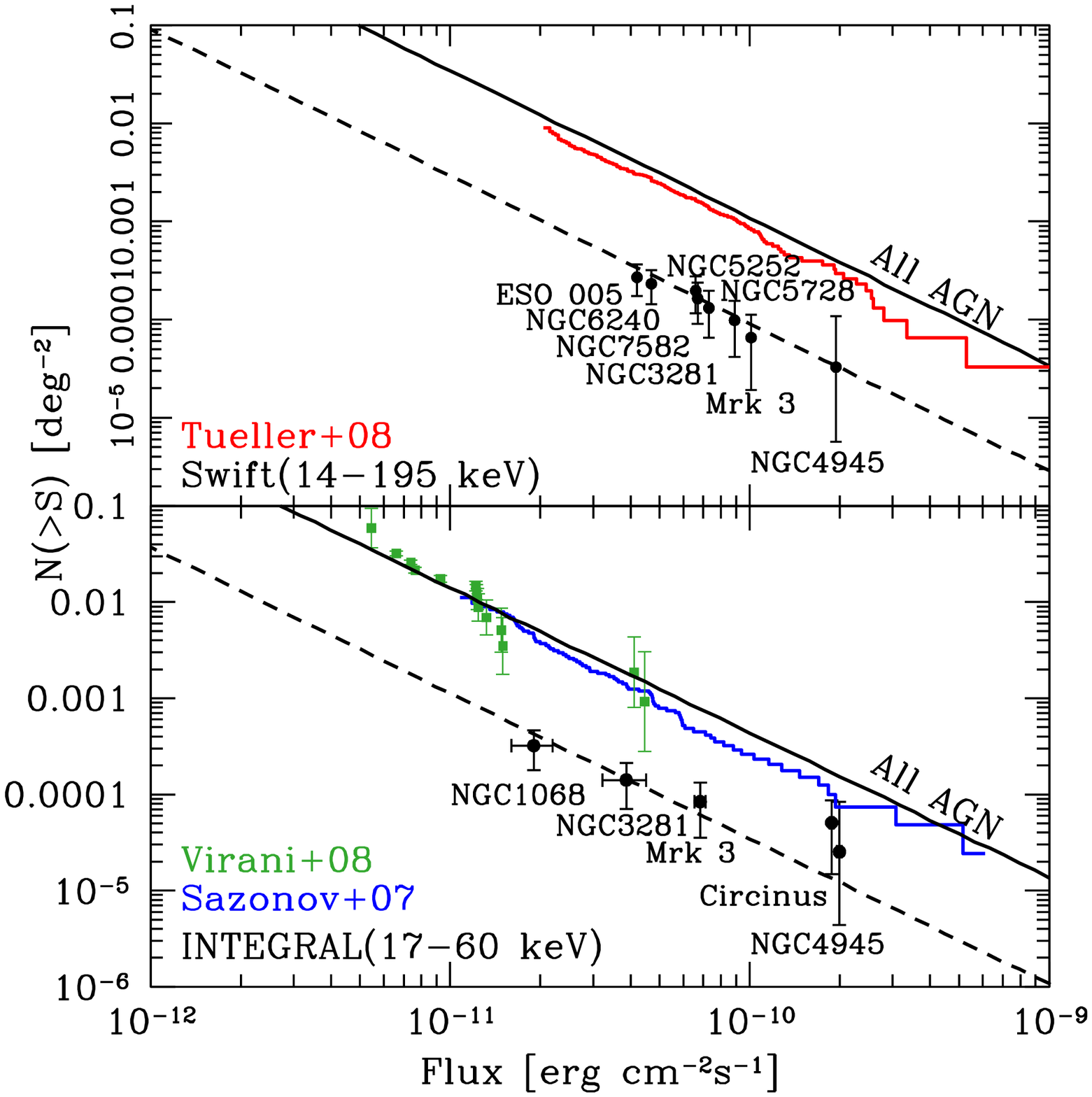}
%\end{center}
\end{minipage} \hfill
%\vspace{-2.5cm}
\begin{minipage}{2.5in}
\caption{LogN-$\log$S distribution for AGN detected at high energies. 
The {\it red line} shows the AGN in the well-defined {\it Swift/BAT} samples 
in the 14-195 keV band \citep{tueller08}, while the {\it bottom panel} shows 
the {\it INTEGRAL} sources \citep{krivonos07} in the 17-60 keV band. {\it Solid squares} 
show the 14 sources detected in the deep 3 Msec {\it INTEGRAL} observations of the 
{\it XMM}-LSS field (S. Virani in prep.). {\it Solid circles} mark the CT AGN detected with 
{\it Swift} ({\it top panel}) and {\it INTEGRAL} ({\it bottom panel}). The {\it black solid
lines} shows the expected AGN $\log$N-$\log$S from the population
synthesis model of \citet{treister05b}, which at these fluxes corresponds to 
a Euclidean distribution. The {\it dashed lines} mark
the Euclidean slope normalized to the number of {\it Swift} and {\it INTEGRAL} 
CT AGN. \label{logn_s}}
\end{minipage}\hfill
\end{ltxfigure}
%\vspace{-2.0cm}

%\section{CT AGN and the X-ray Background}

Since now the number density of CT AGN can be constrained independently, we
can attempt to match the observed spectrum and intensity of the
X-ray background (XRB). In Figure~\ref{xrb}, we show our new fit, which matches the
both the {\it INTEGRAL} and {\it Swift/BAT} observations
at E$>$10~keV and the {\it Chandra} measurements at lower energies (which
are $\sim$30\% higher than the {\it HEAO}-1 A2 observations). These new
data confirmed that the original {\it HEAO}-1 normalization should be
increased by $\sim$30\% and $\sim$10\% at low and high energies
respectively. In contrast, the AGN population synthesis model of
\citet{gilli07} assumed the original {\it HEAO}-1 intensity at all energies,
which translates into a relatively lower contribution from unobscured
sources.  In order to produce the necessary hard spectrum,
\citet{gilli07} had to assume a relatively high number of obscured
sources at high luminosities, i.e., an unusual, inverted dependence of
the obscured fraction of AGN as a function of luminosity
\citep{treister09a}.

\begin{ltxfigure}[h!]
\begin{minipage}{3.6in}
%\begin{center}
\includegraphics[scale=0.4]{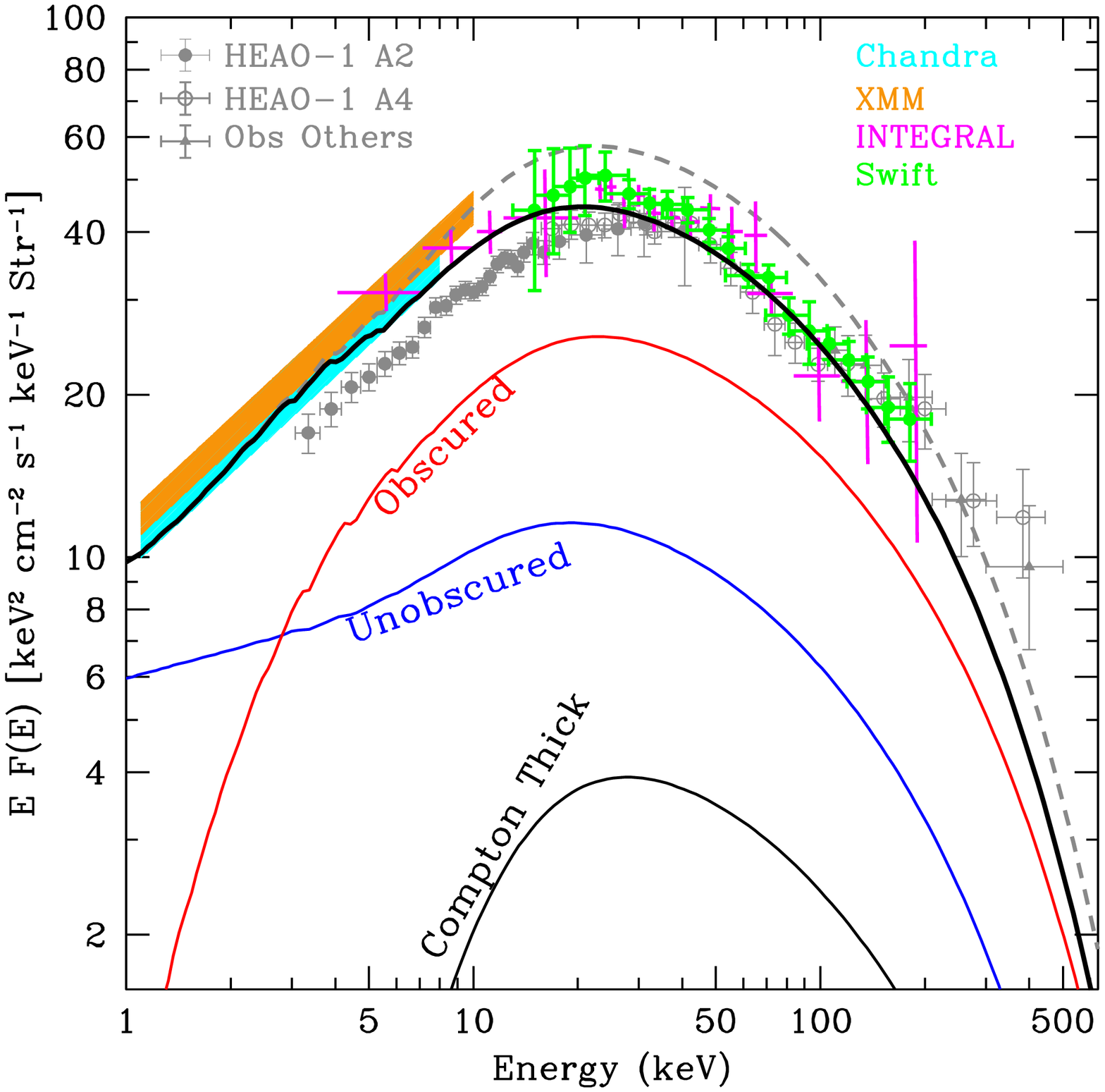}
%\end{center}
\end{minipage}\hfill
%\vspace{-2.7cm}
\begin{minipage}{2.5in}
\caption{Observed spectrum of the extragalactic X-ray background from 
{\it HEAO}-1 \citep{gruber99}, {\it Chandra} \citep{hickox06}, {\it XMM} \citep
{deluca04}, {\it INTEGRAL} \citep{churazov07} and {\it Swift} \citep{ajello08}
data. The {\it dashed gray line} shows the XRB spectrum from the AGN
population synthesis model of \citet{treister05b}, which assumed a 40\% higher
value for the {\it HEAO}-1 XRB normalization. The {\it thick black solid line} shows our 
new population synthesis model for the XRB spectrum; the only change is the number
of CT AGN, which is reduced by a factor of 4 relative to the number in \citet{treister05b}. {\it Red}, {\it blue} 
and {\it thin black} solid lines show the contribution to this model from unobscured,
obscured Compton thin and CT AGN respectively.\label{xrb}}
\end{minipage}\hfill
\vspace{-0.5cm}
\end{ltxfigure}

\section{IR Selection of high-$z$ CT AGN}

Because most of the radiation absorbed at X-ray and UV wavelengths is then re-emitted in the mid- and far-IR, recent 
studies at these energies, mostly taking advantage of {\it Spitzer} observations, have been very successful in finding 
heavily-obscured AGN candidates missed by X-ray selection up to high redshifts. In particular, \citet{fiore08} presented a 
selection method based on the 24~$\mu$m to R band flux ratio and $R$-$K$ color; specifically, $f_{24}$/$f_{R}$$>$1000 
and $R$-$K$$>$4.5 (Vega). According to simulations based on stacking of the  X-ray signal, they estimate the fraction 
of heavily-obscured AGN in this sample to be greater than 80\%.  We applied this selection criteria  to the MIPS-selected 
sources in the ECDF-S to select CT AGN candidates. Of the 7201 24$\mu$m sources detected to a flux limit of 
$\sim$35~$\mu$Jy, 211 ($\sim$3\%) satisfy the $f_{24}$/$f_R$$>$1000 and $R$-$K$$>$4.5 cuts. Of the 651 X-ray 
detected sources in the \citet{virani06} catalog, 18 are found in this region, $\sim$2.8\%, a similar fraction as in the 
general population. 

As can be seen in Fig.~\ref{x_24_corr}, the X-ray to mid-IR ratio for the IR-red excess  sources is about two orders 
of magnitude smaller than the average value for the X-ray-detected sample and for most sources falls in the range 
expected for obscuration of $\sim$5$\times$10$^{24}$ to 10$^{25}$~cm$^{-2}$. The observed ratio for the IR-red 
excess sample is significantly larger than for the sources outside our selection region, even at similar rest-frame 
12~$\mu$m luminosities. The observed X-ray to mid-IR ratio for IR-red excess sources at 
$L_{12\mu m}$$\sim$10$^{45}$~erg~s$^{-1}$ is roughly 3-4 times larger than the values expected from the 
X-ray versus mid-IR luminosity for star-forming galaxies, thus confirming the AGN nature for the vast majority of 
our sample. The space density of CT AGN as a function of redshift inferred from these observations is shown in 
Fig.~\ref{x_24_corr}.

\begin{ltxfigure}[h!]
\begin{center}
\includegraphics[width=0.425\textwidth]{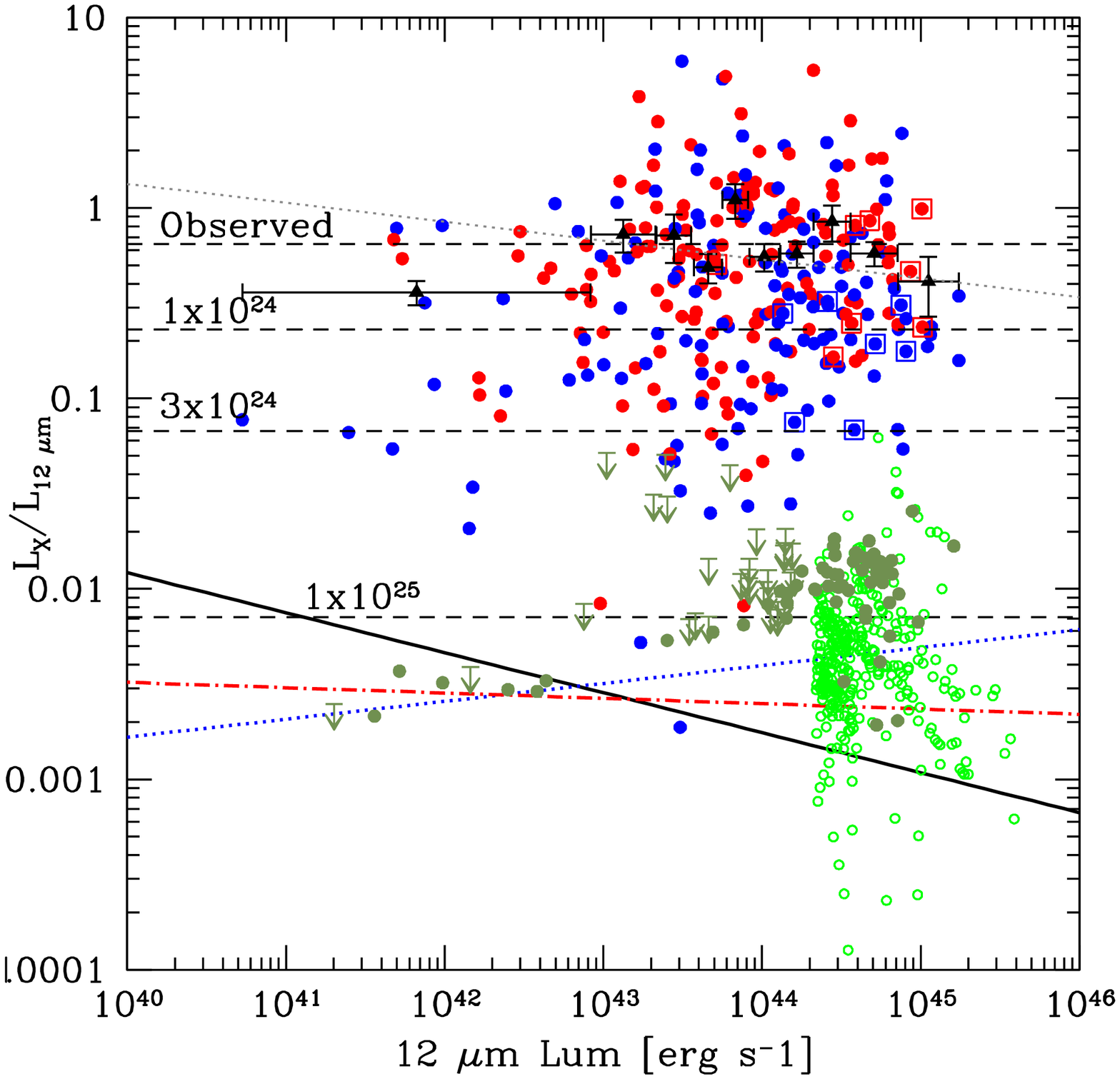}
\includegraphics[width=0.425\textwidth]{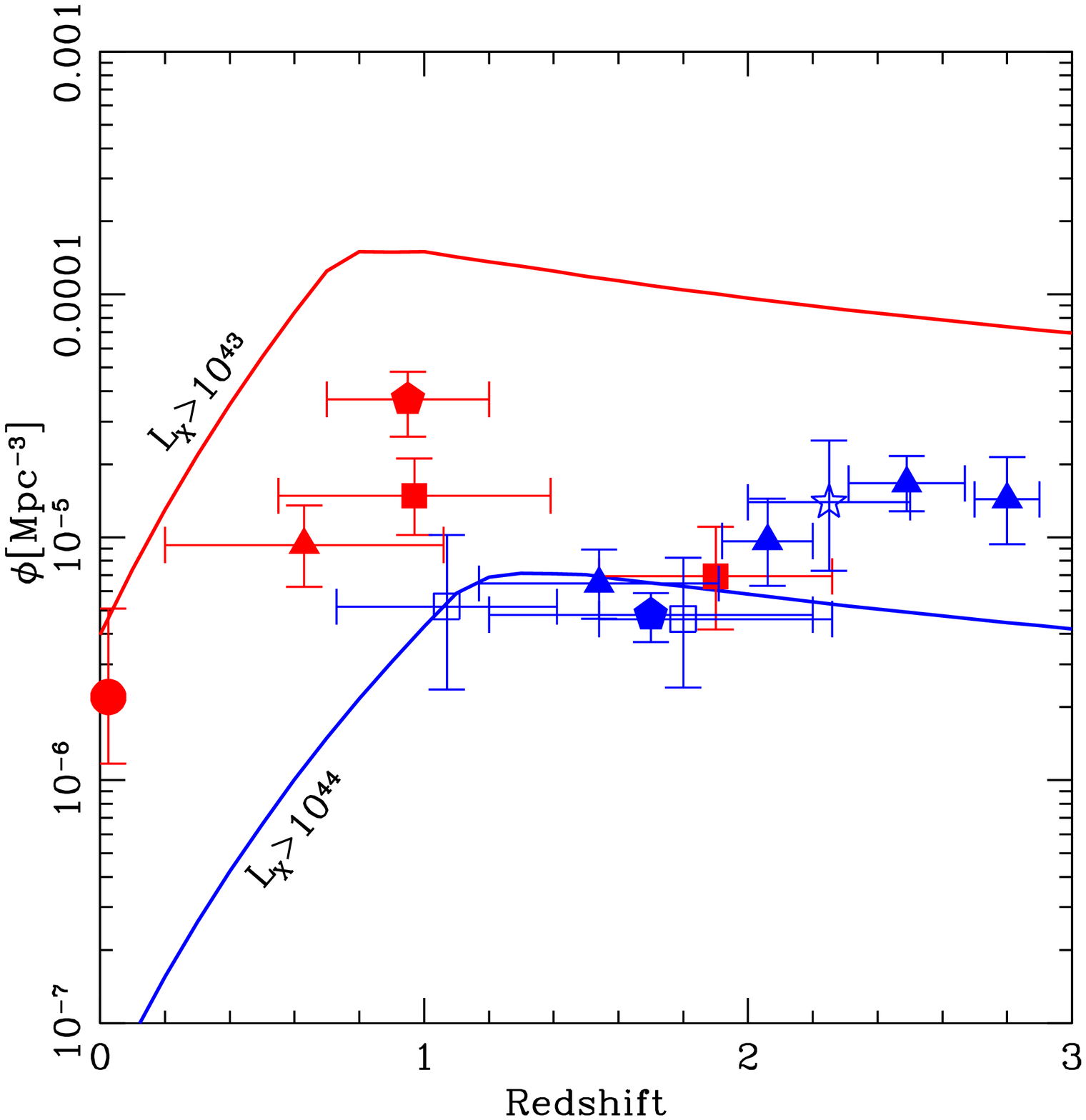}
\end{center}
\vspace{-0.7cm}
\caption{{\it Left panel:} Ratio of hard X-ray to mid-IR versus rest-frame 12~$\mu$m luminosity. {\it Red} 
and {\it blue} circles show the obscured and unobscured X-ray detected sources respectively. {\it Black triangles 
with error bars} show the average ratio for these sources. Circles enclosed by squares identify the X-ray 
detected IR-red excess sources. The {\it dashed line} at a ratio of $\sim$0.6 shows the average value 
of $L_X$/$L_{12~\mu m}$ for all the X-ray sources. The {\it dotted line} shows the relation between
intrinsic X-ray and 12~$\mu$m luminosity for local Seyfert galaxies. {\it Green circles} show the location 
of the most luminous rest-frame 12~$\mu$m X-ray undetected sources outside the selection region. X-ray 
undetected IR-red excess sources are shown by olive circles and upper limits.  {\it Dashed} lines show the 
effects of X-ray obscuration in the observed $L_X$/$L_{12 \mu m}$ ratio. The thick {\it solid, dot-dashed 
and dotted lines} show the expected $L_X/L_{12\mu m}$ for star-forming galaxies, considering different 
recipes to convert star formation rates into X-ray and IR luminosities. {\it Right panel:} Space density of CT 
AGN as a function of redshift. Measurements obtained by \citet{tozzi06,alexander08,fiore09} and this work are
shown by {\it squares}, {\it stars}, {\it pentagons} and {\it triangles} respectively. {\it Solid lines} show the 
expected space density of CT AGN from the luminosity functions of \citet{dellaceca08a}. {\it Red} 
and {\it blue} symbols show  measurements and expectations for $L_X$$>$10$^{43}$ and 
$>$10$^{44}$~erg~s$^{-1}$ sources. These observations indicate a strong increase in the number of 
high-luminosity CT AGN at $z$$>$2.\label{x_24_corr}}
\end{ltxfigure}

%%%%%%%%%%%%%%%%%%%%%%%%%%%%%%%%%%%%%%%%%%%%%%%%
%% BACKMATTER
%%%%%%%%%%%%%%%%%%%%%%%%%%%%%%%%%%%%%%%%%%%%%%%%

\begin{theacknowledgments}
Support for the work of ET and KS was provided by NASA through {\it Chandra}
Postdoctoral Fellowship Award Numbers PF8-90055 and PF9-00069 respectively issued 
by the {\it Chandra} X-ray Observatory Center, which is operated by the Smithsonian
Astrophysical Observatory for and on behalf of NASA.
\end{theacknowledgments}



%\begin{multicols}{2}


%\bibliography{ct_mergers}

%\end{multicols}


\end{document}